\documentclass[a4paper]{article}
\usepackage{fourier}
\usepackage{natbib}
\usepackage[]{graphicx}

\begin{document}

\sloppy
\title{A multispectral and multiscale view of the Sun}

%% Please do not enter footnotes or \inst{}-notes into the optional
%% argument of the author command. The optional argument will go into
%% the header.  If there is only one address the marker \inst{x} may be
%% omitted.

%% Information for the first author.
\author{Thierry Dudok de Wit}

\date{\small{LPC2E, Observatoire des Sciences de l'Univers en r\'egion Centre, 3A avenue de la Recherche Scientifique, 45071 Orl\'eans cedex 2, France} \\
\vspace{5mm}
\small{expanded version of a chapter to appear in: R. Qahwaji, R. Green, and E. Hines (Eds.), \textit{Applied signal and image processing: multidisciplinary advancements}, London, 2010, Bentham Science Publishers.}}

\maketitle                   

\begin{abstract}
The emergence of a new discipline called space weather, which aims at understanding and predicting the impact of solar activity on the terrestrial environment and on technological systems, has led to a growing need for analysing solar images in real time. The rapidly growing volume of solar images, however, makes it increasingly impractical to process them for scientific purposes. This situation has prompted the development of novel processing techniques for doing feature recognition, image tracking, knowledge extraction, etc.  Here we focus on two particular concepts and list some of their applications. The first one is Blind Source Separation (BSS), which has a great potential for condensing the information that is contained in multispectral images. The second one is multiscale (multiresolution, or wavelet) analysis, which is particularly well suited for capturing scale-invariant structures in solar images. 

This chapter provides a brief overview of existing and potential applications to solar images taken in the ultraviolet.
\end{abstract}

\section{Introduction}

The Sun is a world of paradoxes. It is our closest star and yet, distant stars and galaxies have received far more attention as far as data analysis techniques are concerned. Until the dawn of the space age, most solar images were taken in the visible light only, since the terrestrial atmosphere absorbs most other wavelengths. Visible light, however, mostly reveals the photosphere, i.e. the lowest layer of the solar atmosphere, which is relatively featureless apart from the occasional presence of structures such as sunspots. Space-borne telescopes have opened the infrared, the ultraviolet and the X-ray windows, in which the Sun appears much more structured. The vacuum ultraviolet (VUV) range, whose wavelength range extends from 10 to 200 nm, has received considerable attention since it provides deep insight into the highly dynamic and energetic solar atmosphere \citep{aschwanden05b}.

The prime objective of solar image analysis is a better understanding of the complex physical processes that govern the solar atmosphere. The traditional approach consists in observing the Sun simultaneously in different wavelengths and in matching the results obtained by spectroscopic diagnostics with physical models \citep{phillips08}. Indeed, key quantities such as the temperature or the density cannot be directly accessed and so a quantitative picture can only be obtained at the price of time-consuming comparisons with simulations from radiation transfer models, using strong assumptions such as local thermodynamic equilibrium. A key issue is to find new and more empirical means for rapidly inferring pertinent physical properties from such data cubes.

This situation has recently evolved with the emergence of a new discipline called space weather, which aims at understanding and predicting solar variability in order to mitigate its adverse effects on Earth \citep{bothmer06}. Manifestations of solar activity such as flares and interplanetary perturbations indeed influence the terrestrial environment and sometimes cause significant economic losses by affecting satellites, electric power grids, radio communications, satellite orbits, airborne remote sensing and also climate. This new discipline has stimulated the search for new and quicker ways of characterising solar variability. For most users of space weather, empirical quantities that are readily available are valued more than physical quantities whose computation cannot be done in real-time. The sudden need for operational services has stimulated the search for novel multidisciplinary solutions for automated data processing that rely on concepts such as feature recognition, knowledge extraction, machine learning, classification, source separation, etc. \citep{schrijver07b}. The truly multidisciplinary character of this quest is attested by the fact that most of these concepts are also exploited in other chapters of this book.

In most studies of the Sun, the focus has been on the identification and on the characterisation of individual solar features, such as loops \citep{inhester08}, sunspots \citep{gyori02}, prominences \citep{labrosse09} and interplanetary disturbances \citep{robbrecht05}. As the cadence and the size of solar images increases, however, so does the need for extracting metadata and doing data mining. The human eye often remains one of the best expert systems, so tools are also needed to assist humans in visualising multiple images. The Solar Dynamics Observatory satellite, for example, which delivered its first images in April 2010, provides several times per minute $4096 \times 4096$ images in 7 wavelengths in the VUV simultaneously. For such purposes, it is desirable to have techniques that, in addition to extracting specific features, can display 1) multiple wavelengths simultaneously and, 2) multiple scales in a more compact way. Both tasks have not received much attention yet, but will surely become an active field of research in the next decade. 

In this short overview, we shall focus on two particular concepts that are particularly appropriate for handling such tasks; they are multispectral and multiscale analysis. In both cases, potential applications will be emphasized rather than their technical aspects, which can be found in the literature. For recent reviews on feature detection algorithms, see  \citet{perezsuarez10,aschwanden10,zharkova05}. The books by \citet{starck98,starck06} concentrate more on point-like astronomical objects.

\section{A multispectral view of the Sun}

Telescopes that observe the Sun in the VUV are designed to observe preferentially one single and strong spectral line whose emission peaks in a characteristic temperature band. From the simultaneous observation of different spectral lines, one can then build a picture of the temperature layering of the solar atmosphere. More quantitative estimates require a considerable amount of modelling that is still beyond reach. Indeed, many effects such as the integration along the line of sight of optically thin and thick emissions need to be taken into account. 
The solar atmosphere, however, is highly structured by an intense magnetic field. As a consequence, images taken in different wavelengths are often remarkably redundant, as are their time series. This redundancy is manifested by the same location and the similar shape of solar features, as observed in different wavelengths. Redundancy is a useful property for space weather applications because it opens the way for data reduction, i.e. for the extraction of a small set (as compared to the original one) of parameters that describe the salient features of the data. Redundancy also eases the visualisation of multiple wavelengths and is a key ingredient for denoising. A recent and promising framework for dealing with it is blind source separation, which aims at exploiting the coherence in the data to identify their elementary constituents with the aid of the least prior information about them or about their mixing process \citep{choi05,comon10}.

Blind source separation has recently emerged as powerful concept in several areas such as acoustics, data compression, hyperspectral imaging of terrestrial images, etc. \citep{collet10}. Given an instantaneous linear mixture of intensities that are produced by a small set of sources, blind source separation exploits the statistical properties of these sources to differentiate them in an unsupervised manner. This is hallmarked by the cocktail party problem \citep{haykin05}, in which the human ear attempts to isolate one single voice out of a mixture of sources. In the following, we shall see how blind source separation can give new insight into solar multispectral images. Let us first focus on a case study and then discuss the implications.

\subsection{A blind source separation approach}

Figure 1 shows a series of 2D images of the solar limb, taken almost simultaneously by the Coronal Diagnostic Spectrometer (CDS) onboard the SoHO satellite \citep{harrison95}. In this example, the spectrometer was viewing a small region of the Sun and was making 2D raster plots of the intensity of 14 spectral lines. In this particular example, the observation time is negligible as compared to the characteristic evolution time of the solar structures, which can therefore be assumed to be static. CDS offers a similar number of wavelengths as SDO and thus gives a foretaste of what SDO will provide with full Sun images. The intensity of each pixel depends on various plasma parameters along the line of sight, and in particular on the temperature and on the density. A conspicuous feature of these images is their high correlation. There are two reasons for this. First, the temperature response associated with each spectral line is generally wide, and sometimes even multimodal. Second, the same line of sight may capture contributions coming from regions with different temperatures, ranging from about $10^4$ to $10^6$ K. Each image therefore captures contributions associated with a mix of regions (or rather atmospheric layers) that have different temperatures.

\begin{figure}[!htb] 
    \begin{center}
    \includegraphics[width=1.0\textwidth]{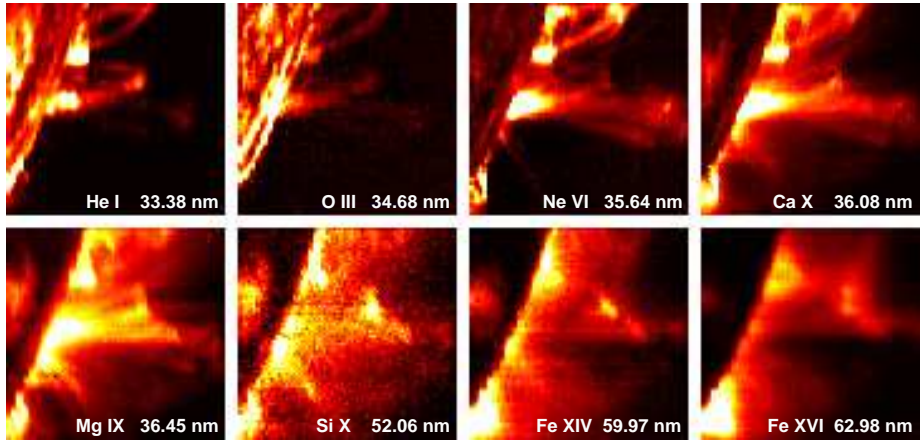}
    \end{center}
    
    \caption{Intensity maps of the same active region at the solar limb taken on March 23, 1998 by the CDS instrument onboard the SoHO spacecraft; 8 emission lines out of 14 are shown. The characteristic temperatures of the lines increases logarithmically from top left ($2 \cdot 10^3$ K) to bottom right ($2.5 \cdot 10^6$ MK). The spectral line and its wavelength are indicated on each image. A linear scale is used for the intensity. \label{fig1}}
\end{figure}

Let us therefore assume that each pixel captures a linear mixture of a large set of pure component spectra, i.e. spectra that are associated with specific emitting regions along the line of sight. Traditionally, individual spectra have been assigned to regions such as coronal holes, the quiet Sun and active regions. The linear mixture hypothesis is reasonably well satisfied in the case of an emitting body like the Sun. For reflecting bodies such as planets, it becomes more debatable \citep{moussaoui08}. The shape of the spectra varies continuously along the line of sight so, a priori, it is not possible to extract the spectra individually. However, since they are partly redundant, one may expect all observations to be described by a small subset of them. That is, the spectral variability should only have few degrees of freedom. There is a wealth of observational evidence in favour of this hypothesis. More than two decades ago, \citep{lean82} had already noticed that the solar spectral variability could adequately be described by 3 different contributions. \citet{feldman08} came to the same conclusion by comparing observations with models while \citet{amblard08} found 3 sources from a statistical analysis of solar spectra. The number 3 therefore seems to be deeply rooted in the spectral characteristics of the solar corona. The exact physical interpretation of this high redundancy has remained elusive. Meanwhile, it provides an ideal framework for doing data reduction by blind source separation.

The CDS spectrometer counts photons, so for each pixel the noise tends to obey a Poisson statistics. We can in principle stabilise the variance by applying the generalised Anscombe transform, which is equivalent here to taking the square root of the intensity, see \citet{starck06}. In the following, we also normalise each image by its mean intensity.

The first step toward blind source separation is the determination of the number of sources. As a first guess we apply principal component analysis or rather the \textit{Singular Value Decomposition} (SVD, \citep{golub00}) to the images and look for the dominant terms. Each image is 85 x 87 pixels in size. We fold the $85 \times 87 \times 14$ data cube into a $7395 \times 14$ matrix by lexicographically ordering each image. By applying the SVD, we implicitly assume that pixel intensities $I(x,\lambda)$ can be expressed as a separable set of orthonormal components that depend either on position $x$ or on wavelength $\lambda$
$$
I(x,\lambda) = \sum_{k=1}^{14} A_k \; f_k(x) \; g_k(\lambda)
$$
with
$$
\langle f_k(x) \; f_l(x) \rangle	= \langle g_k(\lambda) \; g_l(\lambda) \rangle = \left\{
\begin{array}{lll}
0 & \textrm{ if } k \neq l \\
1 & \textrm{ if } k = l 
\end{array} \right.
$$
The weights $A_k$ are by construction positive and they are sorted in decreasing order. The first terms in the sum have the largest weights and thus capture the salient features of the images. The number of components is set by the rank the data matrix, which is here 14.

The distribution of the weights $A_k$ is shown in Figure 2. Their strong ordering confirms the redundancy of the data and suggests that the salient features of the images are expressed by 2 to 5 components only. We obtain similar results without the Anscombe transform, with a weaker ordering of the weights.

\begin{figure}[!htb] 
    \begin{center}
    \includegraphics[width=0.7\textwidth]{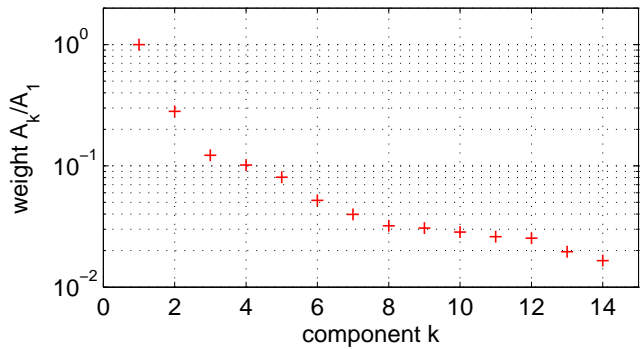}
    \end{center}
    
    \caption{The distribution of the weights obtained by applying the SVD to the 14 images. All weights are normalised to their largest value $A_1$. \label{fig2}}
\end{figure}

The 6 spatial components $f_k(x)$ associated with the six largest weights are shown in Figure 3. These components have no immediate physical for their intensities are non-positive. Note also that there is no sound justification for the orthogonality constraint. 

A more careful analysis suggests that these components also mix different physical features. A more realistic prior is the statistical independence of the components, which is precisely what \textit{Independent Component Analysis} (ICA) does \citep{hyvarinen00}. 

ICA has recently become a popular method for separating sources. This method often brings a significant improvement over the SVD when the probability density of the images is non-Gaussian. Most solar images indeed have a non-Gaussian probability density because they mix features with quite different intensity levels (dark corona vs. bright solar disk). It is therefore worth incorporating this property in the source separation. ICA can be viewed as an inversion of the central limit theorem since it uses the departure from Gaussianity as a lever to improve the discrimination of the sources by assuming that a mixture of random variables is closer to a Gaussian than the individual variables. We illustrate this here by estimating the Kullback-Leibler divergence between the probability density of the images and that of a Gaussian distribution with the same mean and variance. The Kullback-Leibler divergence reads
$$
D(p||q) = \int p(x) \; \log \frac{p(x)}{q(x)} \; \textrm{d}x
$$
where $p(x)$ is the probability density of the image (estimated using a kernel histogram method) and $q(x)$ is a Gaussian density with the same mean and variance. This divergence $D$ is positive; the larger its value, the more $p(x)$ departs from a Gaussian density. We find the divergence of most original images to be between 0.1 and 0.4, whereas that of the 3 main sources estimated by ICA and BPSS (see below) is between 0.3 and 0.8. The source images are as expected more non-Gaussian that the original images.

The sources we obtain by ICA are closer to the expected physical picture than those found by SVD, as shown by \citet{ddw07} for a different data set. However, there is no sound justification for enforcing the independence of the sources; the solar atmosphere is partly transparent and so, for example, in a given active region, different sources may have an intensity peak at the same location. An even more serious objection is the lack of positivity of the sources found.

\begin{figure}[!htb] 
    \begin{center}
    \includegraphics[width=0.8\textwidth]{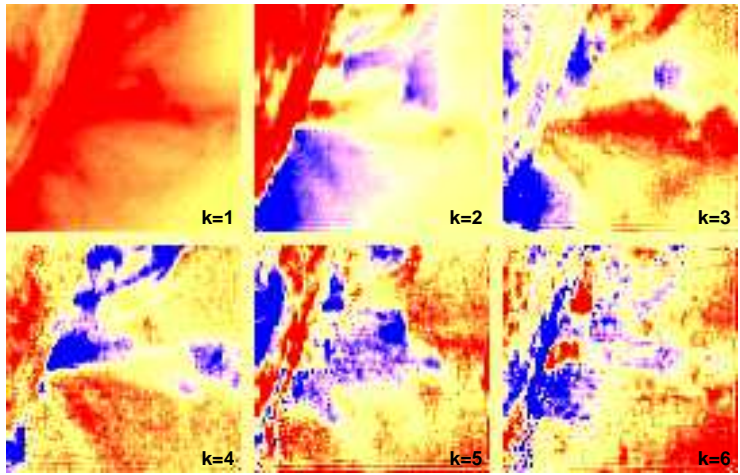}
    \end{center}
    
    \caption{The 6 most significant spatial sources obtained by SVD. The linear vertical scale ranges from the lower 2\% quantile to the upper 98\% quantile, except for the first mode ($k=1$), which is positive. The colour map ranges from blue (negative) to red (positive). \label{fig3}}
\end{figure}

A more realistic prior is the positivity of both the spatial components $f_k(x)$ and their mixing coefficients $g_k(\lambda)$. A natural approach for this is the recent Bayesian Positive Source Separation (BPSS) method, which was developed by \citet{moussaoui06}. BPSS is one among of the several techniques that have recently been developed for doing BSS with positivity constraints. Our motivation for choosing it stems from the Bayesian framework that allows us to incorporate information on the noise and signal statistics. The same method has recently been compared against the ICA in the frame of hyperspectral imaging of Mars \citep{moussaoui08}. 

We assume that 
$$
f(x) = \{f_k(x)\} \qquad g(\lambda) = \{g_k(\lambda)\}
$$
are random matrices whose elements are independent and distributed according to Gamma probability density functions. In the following, the sources $f_k(x)$ have unit norm, as for the SVD. We apply the BPSS to the CDS data after normalising each image to its mean value. No Anscombe transform is used here since we assume each pixel intensity to be a linear mixture of different sources. The question about the number of sources arises again. The root mean square error of the difference between the original data and the reconstructed intensities exhibits a sharp decay as the number of sources increases from one 1 to 3, and then drops more slowly. From this we expect the number of sources to be at least 3.  An inspection of the sources shows that with 4 sources and more, the first 3 ones remain almost unchanged, whereas the other sources are both weaker (i.e. their corresponding mixing coefficients are smaller) and vary with their total number. In other words, the existence of 3 prevalent sources is a robust result, whereas the subsequent sources contribute much less and are not reproducible. A tentative interpretation is that 3 is the right number for obtaining a linear combination, whereas additional terms describe nonlinear effects that do not match some of the hypotheses, such as the positivity of the mixing coefficients. We therefore consider 3 sources in the following.

\begin{figure}[!htb] 
    \begin{center}
    \includegraphics[width=1.0\textwidth]{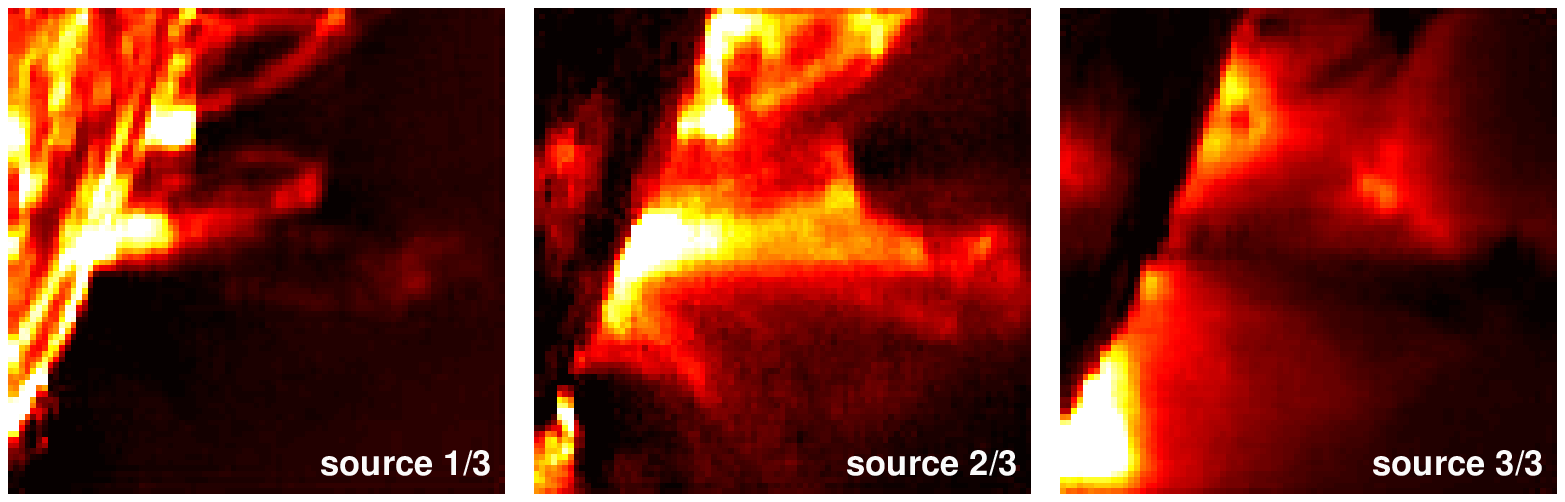}
    \includegraphics[width=0.34\textwidth]{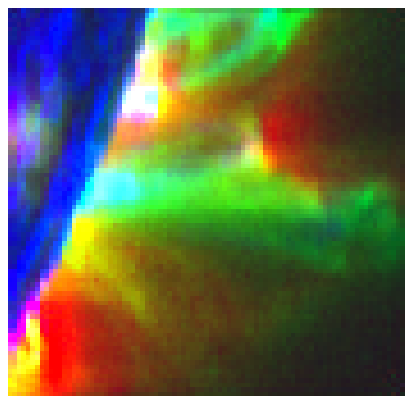}

    \end{center}
    
    \caption{Top row: the 3 sources obtained by BPSS. The linear vertical scale ranges from the lower 2\% quantile to the upper 98\% quantile. All sources are positive definite. Bottom plot: multichannel representation obtained by assigning sources 1, 2 and 3 respectively to the blue, green and red channels. \label{fig4}}
\end{figure}

The 3 sources obtained by BPSS and their mixing coefficients are shown in Figure 4. Our sources are by construction all positive, as are their mixing coefficients. Interestingly, they can be directly linked to specific layers of the solar atmosphere, whereas the interpretation of the sources obtained by SVD or ICA was more difficult. A result of particular interest is the clear temperature ordering in the mixing coefficients. As shown in Figure 5, each source captures emissions that correspond to a specific temperature band. The three sources respectively describe emissions originating from the cool chromosphere (1), from the lower solar corona (2) and from the hot upper corona (3). The cold component is associated with the lowest layers of the solar atmosphere, in which the solar surface comes out as a bright disk. The small loops that stand out against the dark horizon are structured by the solar magnetic field. Particle acceleration processes can locally heat the plasma to several million degrees, leading to the hot diffuse structures that appear in the third source. 

\begin{figure}[!htb] 
    \begin{center}
    \includegraphics[width=0.7\textwidth]{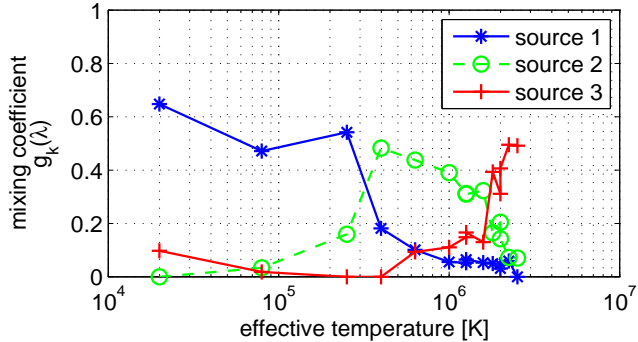}
    \end{center}
    
    \caption{The mixing coefficients associated with the 3 sources displayed in Figure 4. All coefficients are plotted versus the characteristic emission temperature of the corresponding spectral line. \label{fig5}}
\end{figure}

The key result is that all 14 spectral lines, in spite of their differences, can be classified in 3 temperature bands only, whose properties can be inferred from the data without imposing a physical model. These results are robust, in the sense that the same temperature ordering is obtained for other regions or events, provided that all three layers are properly represented in the sample. Preliminary tests with the first VUV images from SDO confirm these results.

\subsection{Other applications}

The concept of blind source separation is better known to the astrophysics community, which has been using it for several years, either for the analysis of multispectral images \citep{nuzillard00} or for the challenging extraction of the cosmic microwave background from images obtained by the Planck mission \citep{kuruoglu10,leach08}. The small number of sources we find in solar VUV images has several practical applications for space weather.

\subsubsection{Temperature maps}

One of the important issues in space weather is the nowcast of the spectrally resolved solar irradiance for the specification of the upper terrestrial atmosphere. Long-term monitoring of the irradiance, however, is difficult because of instrumental constraints. An empirical approach to this problem consists in tracking solar features that emit at different temperatures (e.g. coronal holes, active regions, etc.), assign a characteristic spectrum to each of them, and adding all these contributions to obtain the total irradiance \citep{krivova08}. Segmentation techniques are required for doing this with VUV images. By applying such a segmentation techniques on a few sources only rather than on original solar images, we reduce the computational complexity of the problem. Secondly, by using partly independent source images rather than highly redundant original images we improve the conditioning of the mathematical problem. Thirdly, the segmentation procedure is facilitated if the inputs have a more direct physical interpretation since this allows prior information to be incorporated more easily.

Having 3 sources only is also helpful for condensing all the pertinent information in one single image. We do so by assigning the cold, intermediate and hot sources respectively to the blue, green and red channels. The resulting multichannel representation of the Sun is shown in Fig.\ref{fig4}. This technique, which is commonly used in aerial imagery, allows us to compress all 14 images into one single image, which considerably eases their visual interpretation. We are currently developing real-time three-temperature images of the Sun, which can be used as quicklook plot for locating solar features such as coronal holes.

\subsubsection{Denoising}

Denoising is an interesting but not so common spinoff of blind source separation. Since the salient features of the data are well reproduced by 3 sources only, one may use the latter to reconstruct the observations. The remaining part then mostly consists of incoherent noise. 

One way to investigate the denoising and to qualify the fit is by inspecting the residuals, i.e. the difference between the original image and the image reconstructed from the 3 sources, see Figure 6. Note that the choice of the vertical scale amplifies the residuals; the largest outliers are observed for pixels whose intensity is large as well, so that the relative error actually remains acceptable. Its value generally remains well below 25\% for such pixels.

Residuals are of course particularly interesting for detecting unsuspected features that are not properly described by the model and which may affect one or a few spectral lines locally. In a different application, we detected that way some instrumental and compression artefacts. There aren't any in this particular example.

\begin{figure}[!htb] 
    \begin{center}
    \includegraphics[width=0.98\textwidth]{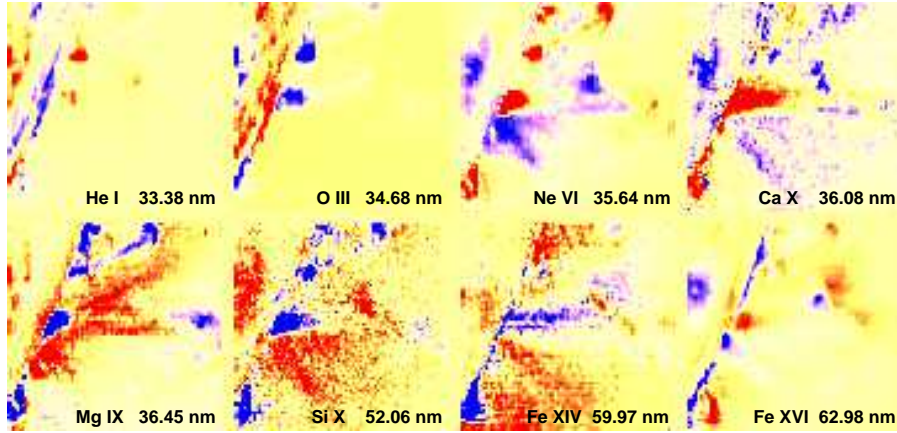}
    \end{center}
    
    \caption{Residuals obtained by removing the contribution of the 3 BPSS sources from the original images, shown in Figure 1. The linear vertical scale ranges from the lower 2\% quantile to the upper 98\% quantile. For each image, this span is about 5 times smaller than the one of the original image. \label{fig6}}
\end{figure}

\subsubsection{Using other physical constraints}

A recurrent problem with blind source separation is the lack of crisp criteria for determining the right number of sources. Although BPSS gives more realistic results than the SVD or ICA, there is still room for improvement. A better separation of the sources requires a quantitatively measurable diversity. Recently, sparsity and morphological diversity have emerged as promising criteria for further improving the separation of sources \citep{bobin08}. Sparsity means that morphologically different features in an image, when projected on a suitable set of basis functions, can be characterised by a small sets of coefficients only, which eases their separation. Another advantage is the possibility to extract more sources than the number of different wavelengths. Solar images exhibit a wide range of different features with characteristic morphologies (e.g. thin loops, diffuse active regions, etc.) and our first results with this approach confirm the relevance of the sparsity concept.

\section{A multiscale view of the Sun}

One of the most fascinating aspects of human vision is its ability to automatically adapt itself to the characteristic scale(s) of features of interest in an image. This has become one of the major challenges in artificial vision and has stimulated the development of multiscale methods for natural images \citep{hyvarinen09}. Solar images, especially when taken in the VUV, provide an excellent example of natural images with a rich blend of multiple scales \citep{delouille05}.
The multiscale analysis of solar images can be performed in many different ways, but wavelet (i.e. multiresolution) techniques are particularly well suited for this \citep{mallat08}. 

Discrete wavelet transforms are widely appreciated for their ability to decompose images into a compact set of coefficients, which can be processed for compression or for filtering and then used for rapid reconstruction of the (filtered) image. The interpretation of these coefficients in terms of physical properties, however, is often difficult. For example, they are not translation invariant and so two images that are shifted by one pixel may have substantially different coefficients. The continuous wavelet transform, which decomposes images into a set of highly redundant coefficients, is much better suited for physical interpretation but its computational burden is considerably higher.

The \textit{\`a trous} (literally, with holes) is one particular algorithm that is popular in astronomical image processing since it shares some of the advantages of the previous methods. This algorithm is relatively fast and yet, the resulting wavelet coefficients have a direct interpretation \citep{mallat08,starck06}.

The decomposition of an image by the \`a trous algorithm is iterative: the original image, which is stored in a rectangular matrix $I$, is convolved with a 2D smoothing kernel $S_d$ whose characteristic scale is $d$, giving a new image $I_d = I*S_d$. The scales are typically dyadic, with $d=0$, 1, 2, 4, 8, \ldots The largest scale must be smaller than the size of the image and $I_0$ is the original image. Other alternatives to the convolution with the smoothing kernel are the pyramidal median transform \citep{starck06}, in which the median over a $(2d+1)\times(2d+1)$ window centred on each pixel is computed. This is particularly appropriate for images that suffer from shot noise. Finally, we build a set of differenced images, 
$$
\{D_0=I_0-I_1, \ D_1=I_1-I_2, \ D_2=I_2-I_4, \ldots, \ D_{N/2}=I_{N/2}-I_N, \ D_N=I_N\} ,
$$
each of which captures structures that have a specific characteristic scale, as in the continuous wavelet transform. Note that we recover the original image simply by adding all differenced ones. 

This decomposition is illustrated in Figure 7 with a solar image taken in the extreme ultraviolet by the recently launched SWAP telescope onboard the PROBA2 satellite \citep{berghmans06}. This image is rich in structures of all sizes. However, because of the variable optical thickness of the solar corona, many structures remain hidden in a haze that hinders their analysis. Thin features outside of the solar disk often also remain unobserved because the disk is so much brighter than the faint corona. For these reasons, it is highly desirable to enhance the image along two different directions
\begin{itemize}
\item For each scale, the contrast between a pixel and its local neighbourhood should be enhanced;
\item For each pixel, the contrast between the different scales should be enhanced.
\end{itemize}

\begin{figure}[!htb] 
    \begin{center}
    \includegraphics[width=0.8\textwidth]{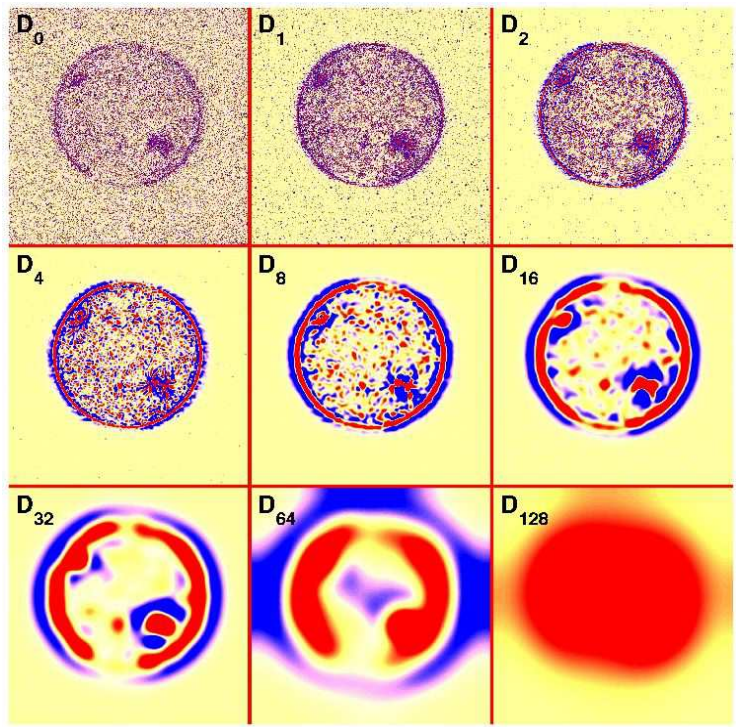}
    \end{center}
    
    \caption{Sequence of differenced images $\{D_0, D_1, \ldots, D_{128}\}$ obtained from a solar image taken in the VUV by the SWAP telescope on Jan. 24, 2010. The original image is shown in Figure 8a and is $1024 \times 1024$ in size. For each image, only intensities ranging from the lower 2\% quantile to the upper 98\% quantile are shown, except for image $D_{128}$, which ranges from 0 to the upper 98\% quantile because all pixels are positive. The colour map ranges from blue (negative) to red (positive). \label{fig7}}
\end{figure}

In this particular example, we process the image in several steps, using the \`a trous algorithm. First the original image is decomposed into a set of differenced images $\{D_0, D_1, \ldots, D_{N}\}$ using a Gaussian smoothing kernel. Next, each differenced image is normalised with respect to its mean absolute intensity, or root mean squared intensity. By doing so, we put all scales on equal footing. The reconstructed image, which is displayed in Figure 8b, already reveals a considerable contrast enhancement. To further enhance the contrast, we normalise for each pixel the wavelet coefficients with respect to their mean absolute intensity, or root mean square intensity. In this last step, the enhancement is done locally only. In contrast to better-known techniques such as histogram equalization, in which the size of the neighbourhood has to be specified using criteria that are often subjective, here the size is adapted automatically to the characteristic size of the locally dominant structure, which is an important asset. This last step considerably enhances structures near the solar limb (the edge of the disk), where their identification is most difficult, see Figure 8c.

In this example, shot noise dominates as soon as one moves away from the limb, so that weak coronal features cannot be followed far into the corona. The image quality can be further improved by adding a filtering stage. This can be done in several ways, either by processing the wavelet coefficients, or by using the discrete wavelet transform and thresholding the wavelet coefficients, see below.

\begin{figure}[!htb] 
    \begin{center}
    \includegraphics[width=1.0\textwidth]{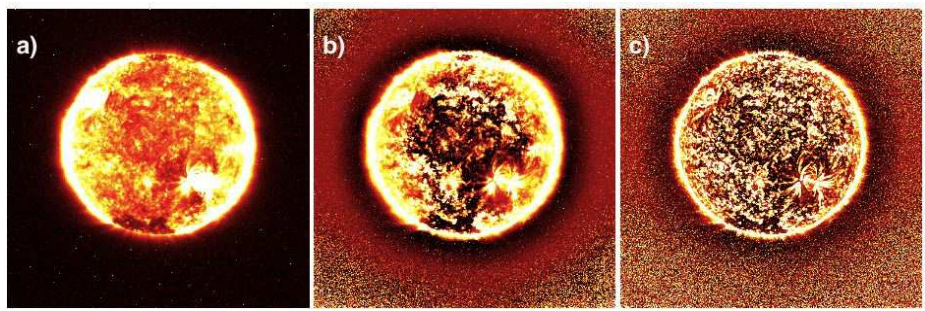}
    \end{center}
    
    \caption{Solar image taken in the VUV by the SWAP telescope. The original (a) and the processed images are shown. In caption (b) only the contrast between scales has been enhanced. In caption (c) contrast enhancement with respect to the local neighbourhood has been included. \label{fig8}}
\end{figure}

\subsection{Other applications}

The multiscale image enhancement can be improved and extended in multiple ways. This is now gradually becoming an active field of investigation in solar physics.

\subsubsection{Feature detection}

The solar feature detection and extraction problem is discussed in detail in a companion article by \citet{perezsuarez10}. Here we consider this problem in the light of multiscale analysis only. The \`a trous algorithm we discussed just before has the advantage of being multi-purpose. When it comes, however, to detecting and extracting structures that have a specific shape, much better performance can be achieved by tailoring the shape of the analysing wavelets to that of the structures of interest. Typical examples are the magnetic loops that often permeate the solar atmosphere and whose conspicuous curved shape requires curved wavelets. Curvelets \citep{candes06} are ideally suited for dealing with such structures, since they have been designed to capture curved shapes. An example is shown in Figure 9, in which the differenced image $D_0$ from Figure 7 has been processed using the discrete curvelet transform. The wavelet coefficients have been computed and only those values exceeding a threshold determined by a preset noise level were retained. The inverse transform should then give an image in which only salient curved features are retained. We find indeed that most of the shot noise has been eliminated that way while arched-like structures in the vicinity of active regions and the solar limb now appear much more evidently. Several successful applications of multiresolution techniques to the detection of solar features have been reported \citep{portierfozzani01,young08,ireland08}. \citet{gallagher10} recently used curvelets to detect coronal mass ejections.

\begin{figure}[!htb] 
    \begin{center}
    \includegraphics[width=0.8\textwidth]{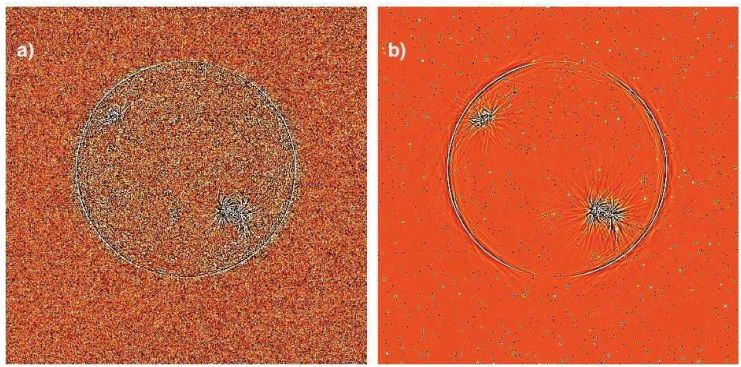}
    \end{center}
    
    \caption{The differenced image $D_0$ from Figure 7 before (a) and after (b) noise reduction using the curvelet transform. The same vertical scale has been used for both images. \label{fig9}}
\end{figure}

It should be stressed that the Haar wavelet, whose popularity stems from its simple shape and its convenient properties, also is one of the worst possible wavelets for analysing solar images. Indeed, the shape of the wavelets should ideally be tailored to the kind of structures one wants to investigate. More exactly, the regularity (or equivalently the number of vanishing moments) in a mother wavelet is directly related to slope of the Fourier power spectral density one can probe. Smooth images have steep spectra, and therefore they are best analysed with high order wavelets \citep{mallat08,abry09}. Haar wavelets are not recommended, unless one is interested in studying discontinuities.

\subsubsection{Denoising}

The problem of denoising images is very similar to that of feature extraction, since both aim at separating desired features from unwanted ones. Multiscale analysis is widely used for that purpose \citep{krim99,mallat08,to09} but applications to solar images are scarce. \citet{stenborg03}, for example, use a wavelet packet approach to extract features and reduce noise. 

Most multiresolution methods are optimal or suboptimal for data that are affected by Gaussian noise. Solar images, however, are often based on photon counting; as a result, the noise characteristics is often a mix of Poisson and Gaussian statistics. In such cases, better results can be achieved by using the Anscombe transform \citep{starck06} to stabilise the variance. In the case of an image that is affected by pure Poisson noise, this transform simply amounts to taking the square root of the pixel intensity.

An interesting issue is the onboard preprocessing of solar images since image compression is often imposed by the limited data flow between satellite and ground. Better performance can be achieved by taking into account the noise statistics in the compression scheme \citep{nicula05}. The next step would be to incorporate the multiscale nature of solar images by doing onboard denoising first. This has not yet been attempted.

\subsubsection{Stereoscopy}

The twin STEREO spacecraft that were launched in 2006 for the first time allowed solar structures to be investigated in 3D. Many tools have been developed for that purpose \citep{wiegelmann09}. Doing tomography with just two vantage points is beyond reach but the 3D stereoscopic reconstruction of contrasted features such as loops is possible \citep{aschwanden05, inhester08}.

Here again, multiresolution methods offer several advantages and are indeed widely used in 3D artificial vision for their capacity of automatically adapting the size of the neighbourhood to that of the salient structures. One typically selects a feature in one image, locates it in the stereoscopic pair, and finally estimates its depth from the disparity (i.e. the displacement). This is the idea behind optical flow \citep{gissot08}, which has been successfully applied to solar images. A multiresolution stage can be added to this by starting the feature matching on the largest scales and refining it down to smaller scales. This not only makes the feature extraction more robust, but it also speeds up the computationally expensive matching procedure.

A powerful alternative to this, which has emerged in the field of artificial vision, is the following: in the continuous wavelet transforms, each image pixel is unfolded into a set of wavelet coefficients that uniquely describe the intensity of that pixel and its local neighbourhood properties such as texture, curvature, etc. If similar features are observed in a stereoscopic pair, then their sets of wavelet coefficients should also closely match. The procedure then goes as follows: take a pixel from one image, and correlate it with pixels in the stereoscopic pair by matching the wavelet coefficients. The two pixels that describe the same structure should be those that have the highest correlation. This has been termed the local correlation function \citep{perrin99} because the correlation is truly done on a pixel per pixel basis. The advantage over the previous methods is the automatic selection of the prevalent scale and the resilience versus distortions; if the same structure does not appear identically in both images (because of the depth), the local correlation function will nevertheless manage to identify it. An application to time series has recently been reported \citep{soucek04} and similar concepts are used in other fields, such as for character recognition.

\subsubsection{Self-similarity}

Wavelets are ideally suited for analysing scale-invariant features in solar images because they are by construction self-similar. Turbulent cascades in the solar atmosphere and the intricate topology of the magnetic field that drives most solar dynamic processes are some among the various reasons for expecting multifractal patterns in solar images. What matters is not the spectral content of the image but rather the interplay between scales.

The multifractal structure of solar images has recently received considerable attention, for quite different reasons. One has to do with the prediction of solar flares from photospheric images and from solar magnetograms (i.e. maps of the photospheric magnetic field). Flares are associated with changes in the topology of the magnetic field, and the idea of actually quantifying the flaring probability from the multifractal structure of the magnetic field has been investigated by several authors \citep{criscuoli07,conlon08,kestener10,mcateer10}. A second motivation for considering the multifractal nature of the solar atmosphere is to shed light on the mechanisms of turbulence and understand the anomalous heating of the corona \citep{georgoulis05}. This has turned out to be difficult because each line of sight integrates emissions originating from various altitudes. A third reason has to do with observational strategies. New solar telescopes have ever increasing spatial and temporal resolutions; as a consequence, the number of photons received per pixel keeps on decreasing (for fixed aperture). The question then arises as to how inhomogeneous the solar emission becomes at small scales \citep{delouille08}. For a multifractal emitting surface, for example, the inhomogeneity should not scale in the same way as for a monofractal surface because the statistical properties differ. This has implications on the definition of the dynamic range of the detector.

Another idea is to use the continuous wavelet transform to investigate the Lipschitz regularity in solar images, i.e. the local ``sharpness'' of discontinuities \citep{mallat08}. VUV images of the solar disk reveal many tiny bright spots, most of which are either transient bright features of the solar corona, or impacts from cosmic rays. The local Lipschitz regularity of these spots depends on their physical origin, thereby giving a means for separating them \citep{hochedez02}.

\section{Conclusion}

In this chapter, we have given a brief tour of solar imaging as seen from a blind source separation and from a multiscale perspective. The recent emergence of space weather has greatly contributed to help introducing concepts that have matured in other fields and that are now just waiting to be applied to the Sun. They are now gaining wider acceptance as it is found that empirical models can be truly complementary to better-known physical models. 
	Blind source separation is particularly useful for providing fast and empirical representations of the temperature distribution in the solar atmosphere. Multiscale methods have even more applications because they are deeply rooted in the scale invariant structure of the solar atmosphere. The best is yet to come as physics-based and empirical models are gradually getting closer to each other, leading to semi-empirical models in which the weaknesses of one can be complemented by the strengths of the other.

\vspace*{1cm}
\subsection*{Acknowledgements}
\small{I would like to thank J. Aboudarham, P.-O. Amblard, F. Auchère, T. Benseghir, G. Cessateur, M. Kretzschmar, J. Lilensten and S. Moussaoui for many stimulating discussions. This study received funding from the European Community's Seventh Framework Programme (FP7/2007-2013) under the grant agreement nr. 218816 (SOTERIA project, www.soteria-space.eu). I thank the PROBA2/SWAP and the SoHO/CDS teams for providing respectively the SWAP and the CDS images. SWAP is a project of the Centre Spatial de Liège and the Royal Observatory of Belgium funded by the Belgian Federal Science Policy Office (BELSPO).}

\bibliographystyle{apalike}
% \bibliography{/Users/ddwit/Documents/Articles_meen/mybib.bib}

\end{document}